\documentclass[reprint,showpacs]{revtex4-1}
\usepackage[T1]{fontenc}
\usepackage[latin9]{inputenc}
\usepackage{amsmath}
\usepackage{amssymb}
\usepackage{graphicx}
\usepackage{esint}

\usepackage{epstopdf}

\makeatletter
\usepackage[colorlinks=true, linkcolor=blue,citecolor=blue,urlcolor=blue]{hyperref}

\makeatother

\begin{document}

\title{Coupling of four-wave mixing and Raman scattering by ground-state atomic coherence}

\author{Micha\l{} Parniak}
\email{michal.parniak@fuw.edu.pl}
\affiliation{Institute of Experimental Physics, Faculty of Physics, University of Warsaw, Pasteura 5, 02-093 Warsaw, Poland}

\author{Adam Leszczy\'nski}
\affiliation{Institute of Experimental Physics, Faculty of Physics, University of Warsaw, Pasteura 5, 02-093 Warsaw, Poland}

\author{Wojciech Wasilewski}

\affiliation{Institute of Experimental Physics, Faculty of Physics, University of Warsaw, Pasteura 5, 02-093 Warsaw, Poland}

\begin{abstract}
We demonstrate coupling of light resonant to transition between two
excited states of rubidium and long-lived ground-state atomic coherence.
In our proof-of-principle experiment a non-linear process of four-wave
mixing is used to achieve light emission
proportional to independently prepared ground-state atomic coherence.
Strong correlations between stimulated Raman scattering light heralding generation
of ground-state coherence and the four-wave mixing signal are measured and shown to survive the storage period, which is promising in terms of quantum memory applications. The process is characterized as a function of laser detunings.
\end{abstract}


\maketitle

\section{Introduction}
Off-resonant Raman scattering is a robust approach to light-atom interfaces.
One of the methods is to induce spontaneous Stokes scattering in which
pairs of photons and collective atomic excitations - a two-mode squeezed
state - are created. These excitations can be stored and later retrieved
in the anti-Stokes process \cite{VanderWal2003,Bashkansky2012}. This approach is
commonly known to be a basic building block of the DLCZ protocol \cite{Duan2001}.

Typically rubidium and cesium have been used as atomic systems in
both warm and cold atomic ensembles \cite{Chrapkiewicz2012,Michelberger2015,Dabrowski2014}.
These systems are coupled to light at near-IR wavelengths,
such as 795 and 780 nm for rubidium D1 and D2 lines. Coupling to new wavelengths holds a promise to greatly extend capabilities of quantum memories. This
can be accomplished by non-linear frequency conversion in the four-wave
mixing (4WM) process using strong resonant non-linearities in atoms
thanks to transitions between excited states. Such processes
have been used to demonstrate frequency conversion in rubidium \cite{Donvalkar2014,Becerra2008,Parniak2015,Akulshin2009a,Khadka2012} or to generate photon pairs in the
cascaded spontaneous 4WM process \cite{Willis2010,Srivathsan2013a,Chaneliere2006}. Multiphoton
processes are also a developing method to interface light and Rydberg
atoms \cite{Kondo2015,Huber2014a,Kolle2012}.

Chaneli\`ere \textit{et al.} \cite{Chaneliere2006} proposed to combine the
processes of Raman scattering and 4WM by first creating photon pairs
in a cascaded spontaneous 4WM in one atomic ensemble and then storing
photons as collective atomic excitations in another cold ensemble,
and an experiment was recently realized \cite{Zhang2016}.
As a result they obtained a two-mode squeezed state of atomic excitations
and telecom photons. Another approach was to frequency-convert light
generated in quantum memory with 4WM \cite{Radnaev2010a}, in order to create a frequency-converted
quantum memory. In all cases one atomic ensemble was used for storage,
and another for frequency conversion or photon generation. 

\begin{figure}
\includegraphics[scale=1.02]{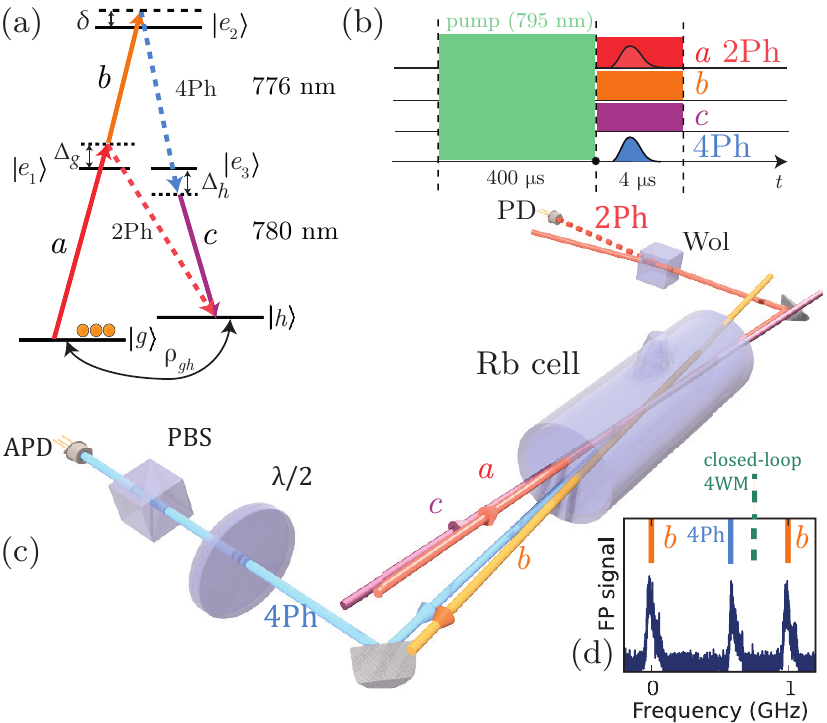}\protect\caption{(a) Configuration of atomic levels and fields we use to realize the
four-photon interface, (b) pulse sequence used in the experiment,
(c) the central part of experimental setup demonstrating phase-matching
geometry and (d) trace of the 1 GHz FSR Fabry-P\'erot interferometer signal showing the frequency of 4Ph field being different from what one would expect from the closed-loop process. Rows in (b) correspond to different beam paths presented
in (c).}
\label{fig:schemat}
\end{figure}

In this paper we realize a Raman-like interface based
on 4WM in warm rubidium vapors driven by ground-state atomic coherence. The process we present may be in principle
used to generate correlated pairs of collective atomic excitations
and photons coupled to transition between two excited states in a
single, four-photon process in a single atomic ensemble. Transition between two excited states corresponds to 776-nm light
as illustrated in Fig.~\hyperref[fig:schemat]{\ref*{fig:schemat}(a)}. As the two intermediate states we use $5\mathrm{P}_{3/2}$
and $5\mathrm{D}_{5/2}$.

The paper is organized as follows. In Sec. \ref{sec:theory} we discuss the principles behind our idea. In Sec. \ref{sec:experimental} we describe the experimental setup and methods we use to verify our findings. Finally, we give the results of our studies of the four-wave mixing interface, namely correlations and statistical properties in Sec. \ref{sec:corr} and detuning dependencies in Sec. \ref{sec:spectral}. We conclude the paper and give prospects for future developments in Sec. \ref{sec:concl}.

\section{General idea}\label{sec:theory}
In our experiment we generate ground-state atomic coherence $\rho_{gh}$
and light denoted by 2Ph in a two-photon stimulated Raman Stokes process, seeded by vacuum fluctuations. The advantage of this approach is the fact that it is a well-established and effective way to prepare atomic-ground state coherence. In particular, it may be used in different regimes, starting from the single-photon and single-excitation spontaneous regime as in the DLCZ protocol \cite{Duan2001,Chrapkiewicz2012}, through the linear gain regime \cite{Raymer1981,Duncan1990} and even in the nonlinear gain-saturation regime \cite{Walmsley1985,Lewenstein1984,Trippenbach1985}. In the two latter cases, macroscopic ground-state coherence is generated \cite{Zhang2014c}. The generated coherence and the number of scattered photons will be highly random but correlated. The atomic coherence is not averaged out to zero due to atomic motion, since the buffer gas makes the motion diffusive \cite{Raymer2004}. Moreover, the generated Raman field remains coherent with the driving field, so phase fluctuations of the driving field do not disturb the process \cite{Wu2010}. In particular, the generated macroscopic ground-state coherence may be probed \cite{Chen2010}, read-out \cite{VanderWal2003}, or may enhance further stimulated Raman process \cite{Yuan2010}.

In this experiment, we observe concurrent generation of 776-nm light denoted by 4Ph in
a four-photon process analogous to stimulated Raman scattering driven by
ground-state coherence. It does not occur spontaneously in the macroscopic
regime due to small gain. However, with macroscopic $\rho_{gh}$ generated in the two-photon stimulated Raman process, the
driving fields $a$, $b$ and $c$ couple ground-state atomic coherence
to the weak optical 4Ph field. In other words, the 4Ph process is stimulated by pre-existing atomic coherence. In the leading order in drive beam
fields Rabi frequencies $\Omega_{i}$ the atomic polarization resulting
in emission of 4Ph signal field is:

\begin{equation}
\mathbf{P}_\mathrm{4Ph}=-n\mathbf{d}_{e_{3}e_{2}}\rho_{gh}\frac{{\Omega_{a}\Omega_{b}\Omega_{c}^{*}}}{4\Delta_{g}\delta\Delta_{h}},
\label{eq:P2ndrho}
\end{equation}

where $n$ is the atom number density and $\mathbf{d}_{ij}$ is the
dipole moments of respective transition. From this formula it follows
which detunings play role.

For the experimental observation \textit{a priori} knowledge of polarization
properties is crucial. To find it, we add contributions from
all possible paths through intermediate hyperfine states. Since the
detunings from intermediate $5\mathrm{P}_{3/2}$ state are much larger
than respective hyperfine splittings we ignored the latter. Same approximation
is adopted for the highest excited state $5\mathrm{D}_{5/2}$.
Even though respective detuning $\delta$ it is of the order of several
MHz, similar to the hyperfine splitting of the highest excited state
$5\mathrm{D}_{5/2}$, the hyperfine structure is completely unresolved
due to significant pressure broadening \cite{Zameroski2014} in 0.5~torr krypton as buffer gas we use. Consequently, we may omit any detuning
dependance and calculate the unnormalized polarization vector $\boldsymbol{\epsilon}$
of the signal light using path-averaging formalism we developed in
\cite{Parniak2015}, by decalling the definitions of Rabi frequencies
in Eq.~\hyperref[eq:P2ndrho]{(\ref*{eq:P2ndrho})}:

\begin{equation}
\boldsymbol{\epsilon}_\mathrm{4Ph}\propto\sum_{F,\:m_{F}}\mathbf{d}_{e_{3}e_{2}}(\mathbf{E}_{a}\cdot\mathbf{d}_{ge_{1}}\mathbf{E}_{b}\cdot\mathbf{d}_{e_{1}e_{2}}\mathbf{E}_{c}^{*}\cdot\mathbf{d}_{e_{3}h}^{*}),
\label{eq:polaryzacje}
\end{equation}

where $\mathbf{E}_{i}$ are electric fields of respective beams $i$.
Summation is carried over all possible magnetic sublevels ($F_{i}$,
$m_{F_{i}}$) of all intermediate states $|e_{1}\rangle$, $|e_{2}\rangle$
and $|e_{3}\rangle$. 
\section{Experimental Methods}\label{sec:experimental}
The experimental setup is built around the rubidium-87 vapor cell
heated to 100$^{\circ}$C, corresponding to atom number density $n=7.5\times10^{12}\ \mathrm{cm^{-3}}$ and optical density of 1600 at D2 line for optically pumped ensemble. For the two lowest, long-lived states,
we use two hyperfine components of rubidium ground-state manifold,
namely $|5\mathrm{S}_{1/2},F=1\rangle=|g\rangle$ and $|5\mathrm{S}_{1/2},F=2\rangle=|h\rangle$.
For the $|e_{1}\rangle$ and $|e_{3}\rangle$ states we take hyperfine
components of the $5\mathrm{P}_{3/2}$ manifold, and for the highest
excited state $|e_{2}\rangle$ we take the $5\mathrm{D}_{5/2}$ manifold.
Atoms are initially pumped into $|g\rangle$ state using 400 $\mu$s
optical pumping pulse (at 795 nm). Next, three square-shaped driving pulses of 4 $\mu$s duration each
are applied simultaneously, as shown in Fig.~\hyperref[fig:schemat]{\ref*{fig:schemat}(b)}. Inside
the cell, beams are collimated and have diameters of 400~$\mu$m,
being well-overlapped over a 2-cm-long cylindrical region. They intersect
at a small angle of 14 mrad, with 780-nm beams nearly counter-propagating
with respect to the 776-nm beams, as presented in Fig.~\hyperref[fig:schemat]{\ref*{fig:schemat}(c)}.
Powers of driving beams $a$, $b$ and $c$ are 10 mW, 45 mW and 8
mW, respectively. 

The 780-nm two-photon Raman signal 2Ph co-propagates with the driving
field $a$. It is separated using a Wollaston polarizer (Wol) and
detected on a fast photodiode (PD), with $10^4$ signal to driving field leakage ratio. The four-photon signal 4Ph, is
emitted in the direction given by the phase matching condition $\mathbf{k}_{a}+\mathbf{k}_{b}=\mathbf{k}_{\mathrm{{4Ph}}}+\mathbf{k}_{c}+\mathbf{K}$.  The wavevector $\mathbf{K}$ of spin-wave generated in the 2Ph process Raman scattering equals $\mathbf{K}=\mathbf{k}_a-\mathbf{k}_{\mathrm{2Ph}}$. We note that both longitidual and transverse components of the spin-wave $\mathbf{K}$ are much smaller than that of light wavevectors, and consequently it has only an insignificant effect on the 4Ph signal emission direction. This particular, nearly co-propagating geometry allows us to couple the same spin-wave of wavevector $\mathbf{K}$ to both the 2Ph and 4Ph processes.
In addition to the desired 4Ph signal, 776-nm light coming from the
closed-loop process in which field $c$ couples level $|e_{3}\rangle$
directly to $|g\rangle$ is emitted in the same direction \cite{Parniak2015}
and at a frequency differing by only 6.8~GHz. However,
when driving fields $a$, $b$ and $c$ are $x$-, $y$- and $y$-
polarized, respectively, the 4Ph signal light is $x$-polarized, while
the closed-loop 4WM light is $y$-polarized, where $x\perp y$. This arrangement enables
filtering out the 4Ph signal from both stray 776-nm driving light
and the closed-loop 4WM light with a polarizing beam-splitter (PBS, $10^2$ extinction).
To suppress residual drive laser background at 780 nm 4Ph signal goes through
an interference filter (transmission of 80$\%$ for 776 nm and $10^{-3}$ for 780 nm) and is detected by an avalanche photodiode
(APD). We were able to obtain signal to background ratio of up to $10^2$. 

By rotating the half-wave
plate ($\lambda/2$) we can easily switch between observing 4Ph
and closed-loop signal. For detunings optimal for the 4Ph process, we register less than 10 nW of the closed-loop 4WM light.  The two signals display different temporal characteristics,
also only 4Ph is correlated with the 2Ph light. 

We verify the frequencies of 4Ph and closed-loop light using
a scanning Fabry-P\'erot confocal interferometer of 1 GHz free spactral range (FSR) inserted in the 4Ph beam path. Fig.~\hyperref[fig:schemat]{\ref*{fig:schemat}(d)} show the trace obtained by scanning the interferometer for detuning difference $\Delta_g+\Delta_h=2\pi\times3.43$ GHz. Leakage of driving field $b$, induced by slight misalignement, is used as frequency reference. The middle peak corresponds to the 4Ph signal, with a solid line indicating expected frequency. Dashed line corresponds the frequency one would expect from the closed-loop signal.

\section{Results}
\subsection{Statistics and Correlations}\label{sec:corr}
\begin{figure}[b]
\begin{centering}
\includegraphics[scale=1.2]{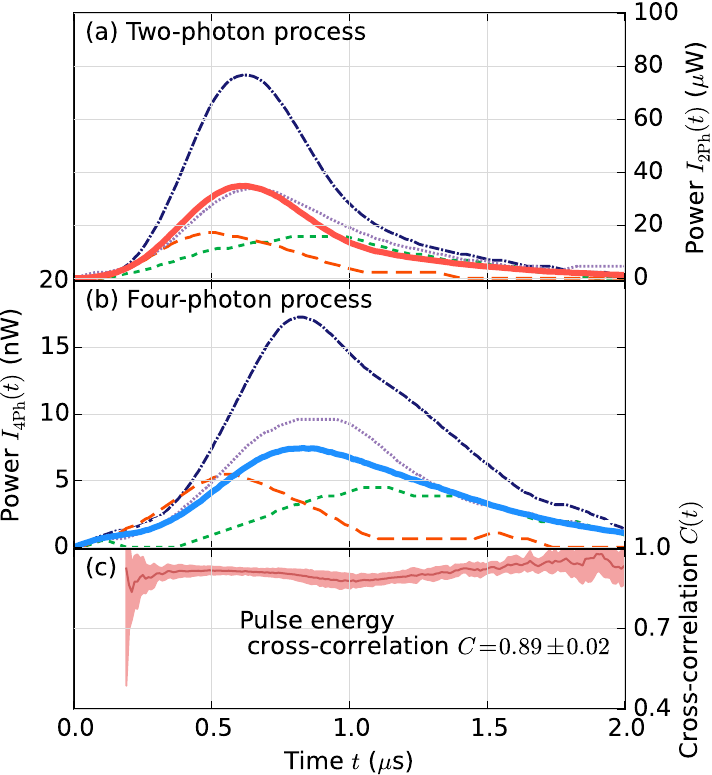}
\par\end{centering}

\protect\caption{(a - 2Ph, b - 4Ph) Averaged signal intensities (solid lines) with several
single realizations (dashed lines), denoted by same colors in (a)
and (b), demonstrating visible strong correlations and (c) calculated
cross-correlation $C(t)$ between signals of four-photon (4Ph) and
the two-photon processes (2Ph). }
\label{fig:corr}
\end{figure}

\begin{figure}[b]
\begin{centering}
\includegraphics[scale=0.7]{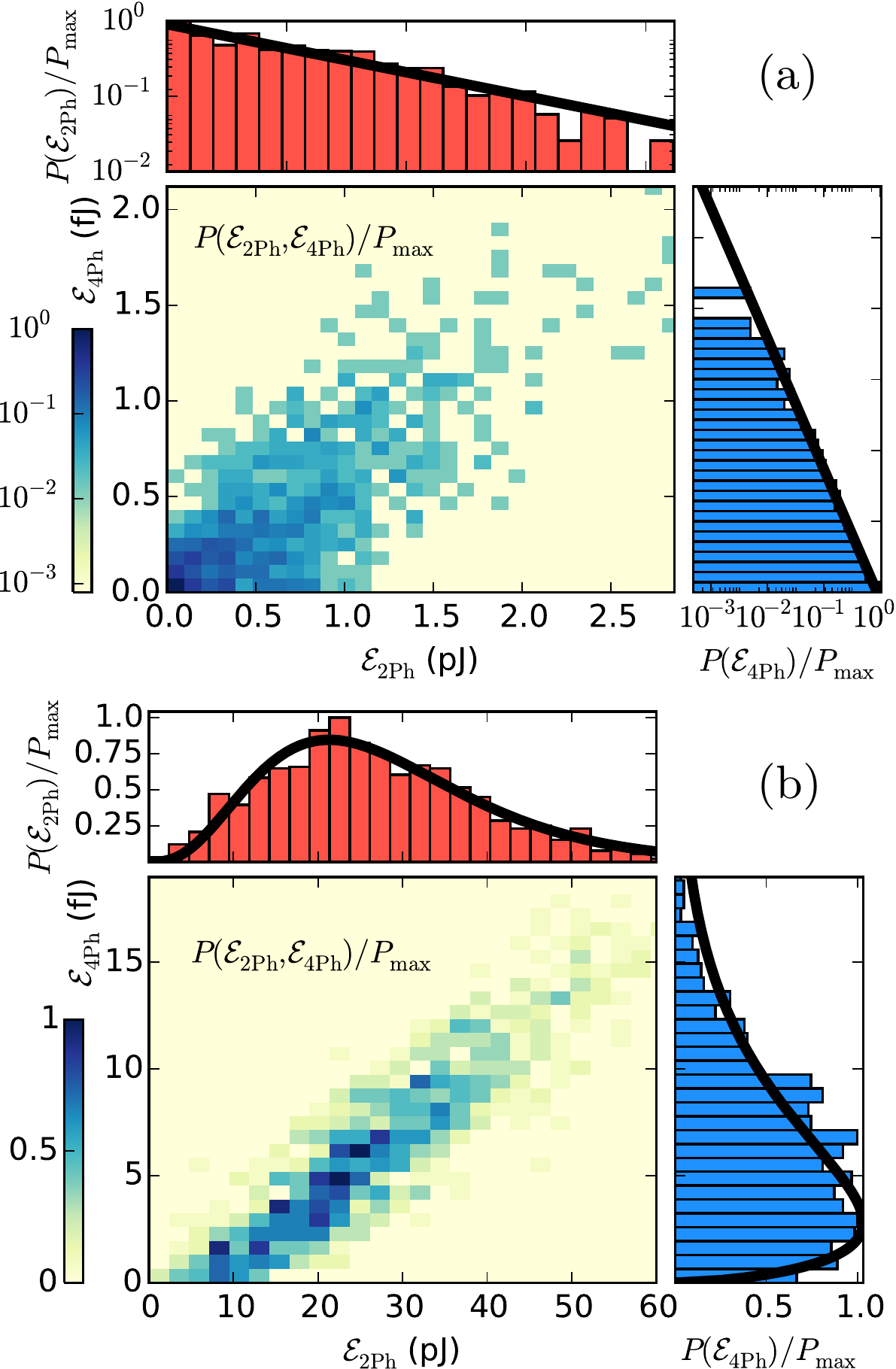}
\par\end{centering}

\protect\caption{Joint statistics $P(\mathcal{E}_\mathrm{2Ph},\mathcal{E}_\mathrm{4Ph})$ together with marginal distributions of registered pulse energies for (a) short driving time of 200 ns, yield nearly single-mode thermal statistics (note the logarythimc scale in this plot), and (b) longer driving time of 4 $\mu$s with observable pulse energy stabilization well described by multimode thermal statistics with mode number $\mathcal{M}$ of 4.1 for 2Ph pulses and 1.6 for 4Ph pulses. Solid curves correspond to fitted thermal distributions.}
\label{fig:corrmap}
\end{figure}

In our experiment we remain in the macroscopic scattered light intensity
regime. When strong Raman driving field is present, atoms are transferred
to $|h\rangle$ simultaneously with scattering of
2Ph photons and buildup of large atomic coherence $\rho_{gh}$.
The temporal shape of the 2Ph pulse is an exponential only at the
very beginning, which we observe in Fig.~\hyperref[fig:corr]{\ref*{fig:corr}(a)}. Not only pulse energies, but also the shapes fluctuate significantly from shot to shot, as the process is
seeded by vacuum fluctuations \cite{Raymer1989}. However, the
4Ph pulse, presented in Fig.~\hyperref[fig:corr]{\ref*{fig:corr}(b)}, nearly exactly follows
the 2Ph pulse. We calculate temporal correlations between the 2Ph
signal, which is known to be proportional to the ground-state atomic
coherence $\rho_{gh}$, and the 4Ph signal. Normalized intensity
correlation at time $t$ between the 2Ph signal $I_{2\mathrm{Ph}}(t)$
and the 4Ph signal $I_{4\mathrm{{Ph}}}(t)$ is calculated according
to the formula $C(t)=\langle\Delta I_{\mathrm{{2Ph}}}(t)\Delta I_{\mathrm{{4Ph}}}(t)\rangle/\sqrt{\langle\Delta I_{\mathrm{{2Ph}}}^{2}(t)\rangle\langle\Delta I_{\mathrm{{4Ph}}}^{2}(t)\rangle}$
by averaging over 500 realizations, where the standard deviations
$\langle\Delta I_{\mathrm{{2Ph}}}^{2}(t)\rangle$ and $\langle\Delta I_{\mathrm{{4Ph}}}^{2}(t)\rangle$
are corrected for electronic noise. Figure \hyperref[fig:corr]{\ref*{fig:corr}(c)} presents
the cross-correlation $C(t)$. We observe that correlations are high
during the entire process, which proves that at any time both processes
interact with the same atomic coherence $\rho_{gh}$, similarly as in some previous works in $\Lambda$-level configurations \cite{Chen2010,Yang2012b}.
In particular, we are able to measure high correlation >0.9 at the very
beginning of the pulses, were light intensities are low. This regime
is quite promising for further quantum applications. To estimate the
uncertainty of calculated correlations, we divided data into 10 equal
sets of 50 repetitions and calculated the correlation inside each
set to finally obtain the uncertainty by calculating standard
deviation of results from all sets. 

Next, we study statistics and correlations of pulse energies in detail. We consider short and long pulse duration regimes. Figure \hyperref[fig:corrmap]{\ref*{fig:corrmap}(a)} corresponds to short driving time of 200 ns. In this regime light generated in both 2Ph and 4Ph processes is well-characterized by single-mode thermal energy distribution $P(\mathcal{E})=\langle\mathcal{E}\rangle^{-1} \exp(-\mathcal{E}/\langle\mathcal{E}\rangle)$ \cite{Raymer1982}, where $\mathcal{E}$ is the total scattered light energy in a single realization. This observation shows that we excite only a single transverse-spatial mode, as intended by using an adequately small size of Raman driving beam $a$ \cite{Raymer1985a,Duncan1990}. Thermal distribution yields very high pulse energy fluctuations (namely, mean energy is equal to standard deviation), that are due to vacuum fluctuations of electromagnetic field and quantum fluctuations of atomic state \cite{Walmsley1985,Walmsley1983}. Still, we observe that energies of 2Ph and 4Ph pulses are highly correlated, which is demonstrated by  joint statistics $P(\mathcal{E}_\mathrm{2Ph},\mathcal{E}_\mathrm{4Ph})$. The residual spread is mainly due to detection noise of both signals.

In the second regime [Fig. \hyperref[fig:corrmap]{\ref*{fig:corrmap}(b)}] the driving pulses of 4~$\mu$s are longer than in the previous scheme. The relative fluctuations become smaller due to gain saturation. We found the marginal statistics to be well described by multimode thermal distributions (with number of modes $\mathcal{M}$) given by 
\begin{equation}
P(\mathcal{E})=\frac{\mathcal{M}}{\langle\mathcal{E}\rangle\Gamma(\mathcal{M})}\left(\frac{\mathcal{E}\mathcal{M}}{\langle\mathcal{E}\rangle}\right)^{\mathcal{M}-1}\exp(-\mathcal{E}\mathcal{M}/\langle\mathcal{E}\rangle).
\end{equation}
The joint statistics $P(\mathcal{E}_\mathrm{2Ph},\mathcal{E}_\mathrm{4Ph})$ demonstrate clear correlations, which are here less influenced by detection noise than previously, as the pulse energies are higher.

\begin{figure}[b]
\begin{centering}
\includegraphics[scale=0.75]{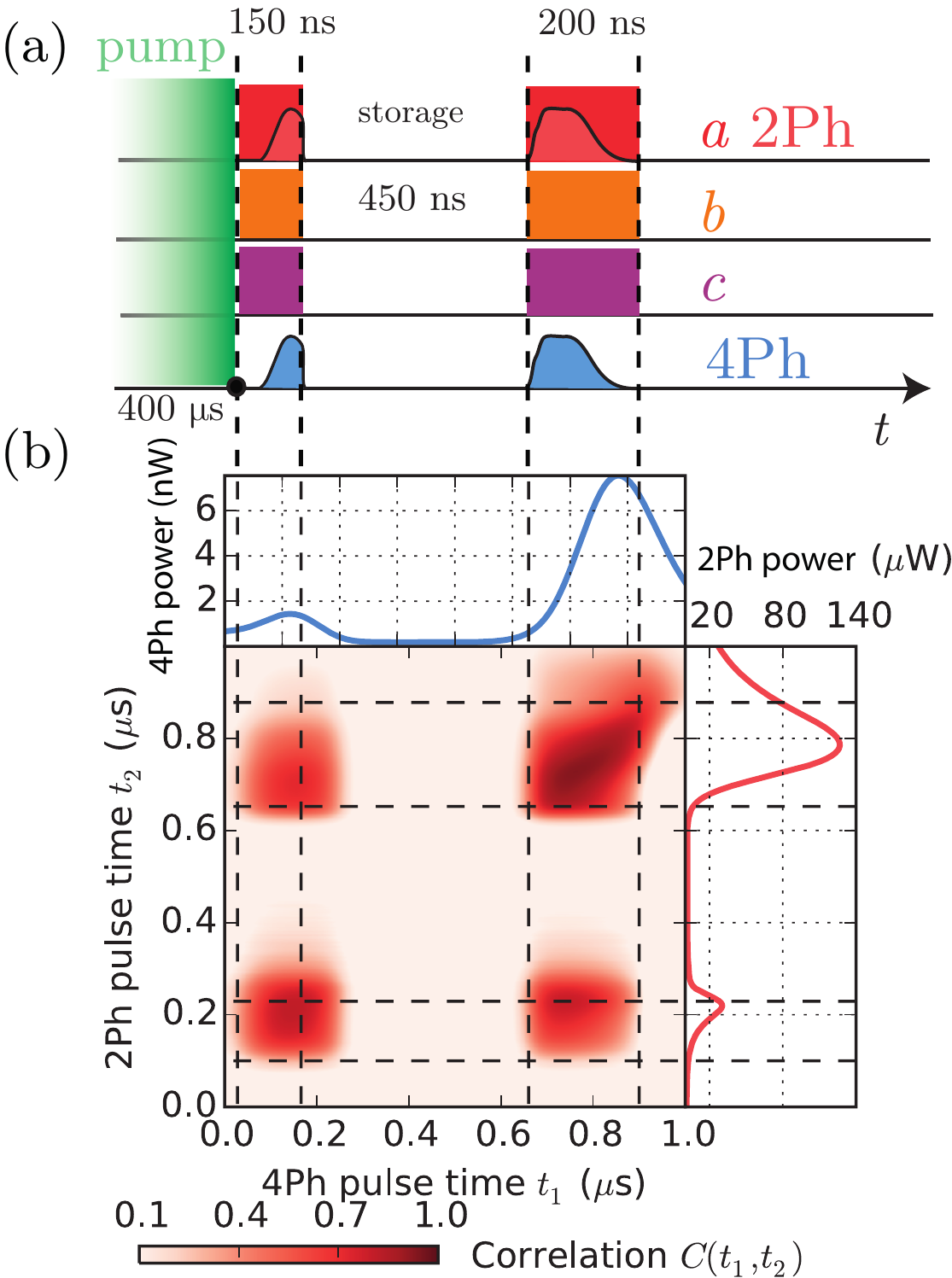}
\par\end{centering}
\protect\caption{(a) Scheme of experiment to demonstrate storage of ground-state atomic coherence for $\tau=450$ ns. (b) Measured two-point correlation $C(t_1,t_2)$ between 2Ph and 4Ph signals with average registered signal powers. Off-diagonal elements of the correlation map demonstrate that light fields interact with the same atomic coherence before and after the time delay. }
\label{fig:corrmaptemporal}
\end{figure}

Finally, we check that correlations are indeed mediated by ground-state atomic coherence by interrupting the scattering process for a dark period of $\tau=450$ ns. This is proved by strong correlations between the intensities of light scattered before and after the dark period, observed in both 2Ph and 4Ph processes. 

After the atoms are optically pumped as in the original scheme, we drive the processes with 150 ns pulses of the three driving fields $a$, $b$ and $c$. After a dark period of $\tau=450$ ns, we drive the process for another 200 ns. The coherence $\rho_{gh}$ generated in the first stage induces optical polarization at both the 4Ph field frequency (as in Eq. \ref{eq:P2ndrho}) and 2Ph field frequency:
\begin{equation}
\mathbf{P}_\mathrm{2Ph}=-n\mathbf{d}_{e_h e_1}\rho_{gh}\frac{\Omega_a}{\Delta_g},
\label{eq:2ph}
\end{equation}
resulting in stimulated Raman emission. The full two-point correlation map presented in Fig.~\hyperref[fig:corrmaptemporal]{\ref*{fig:corrmaptemporal}(b)} is calculated as $C(t_1,t_2)=\langle\Delta I_{\mathrm{{4Ph}}}(t_1)\Delta I_{\mathrm{{2Ph}}}(t_2)\rangle/\sqrt{\langle\Delta I_{\mathrm{{4Ph}}}^{2}(t_1)\rangle\langle\Delta I_{\mathrm{{2Ph}}}^{2}(t_2)\rangle}$. Apart from the diagonal correlated areas at $t_1\approx t_2 \approx 100$ ns and 800 ns, we observe the anti-diagonal terms corresponding to correlations between the two pulses. Due to spontaneous emission and collisional dephasing all the excited atomic states decay quickly, with their lifetime limited to 20 ns. This proves that we store information in the ground-state atomic coherence. Storage time $\tau$ is limited mainly by atomic motion. In fact, the stored ground-state coherence and in turn correlations decay in multiple ways, e.g. by diffusive spatial spread of atomic coherence \cite{Chrapkiewicz2014b,Parniak2014} and by influx of optically pumped atoms into the interaction region.
Finally, we mention that very similar results are obtained if the system is driven by field $a$ only in the first stage of the sequence and in turn only the 2Ph field and atomic coherence are generated. In the second stage of the sequence, where all driving field are present, we observe that both the 2Ph and 4Ph signals are correlated with the 2Ph signal emitted in the first stage.

Agreement of the above measured statistics with theoretical predictions and multiple previous experiments proves that the correlations do arise from vacuum fluctuations. With driving power stable to within less than 1\% and laser frequency stable to within 1 MHz, these external contributions to fluctuations can be neglected. Large magnitude of fluctuations, with nearly perfect correlations between 2Ph and 4Ph signals together with demonstrated storage of correlated signals allow to reject phase-noise to amplitude-noise conversion as a source of correlations \cite{Cruz2006}.

All of the above results were measured for $\Delta_g/2\pi=1000$ MHz, $\Delta_h/2\pi=1200$ MHz and $\delta/2\pi=-50$ MHz.

\subsection{Detuning dependance}\label{sec:spectral}

\begin{figure}
\includegraphics[scale=1.03]{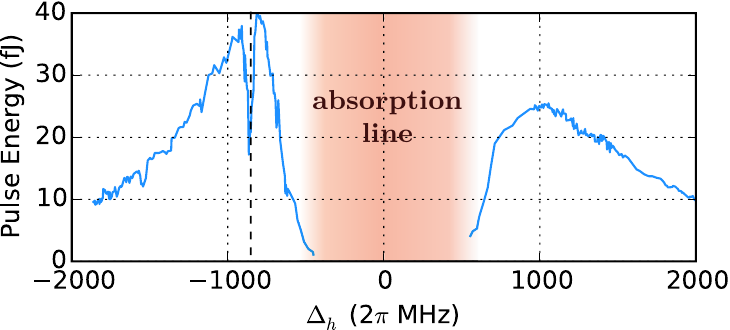}\protect\caption{(a) Averaged (over 500 realizations) pulse shapes
for the intensities of the 2Ph and the 4Ph for a set of single-photon
detunings $\Delta_{g}$ for constant $\delta/2\pi=-50$
MHz and $\Delta_h/2\pi=1500$ MHz, (b) pulses full-width at half-maxima (FWHM), (c) their energies and (d) pulse energy correlation between the two processes as
a function of field $a$ single photon detuning $\Delta_{g}$. Subsequent
plots of 2Ph (4Ph) signals in (a) are shifted by 120 $\mu$W (40 nW).}
\label{fig:shapes}
\end{figure}

Now we switch to verifying properties of 4Ph signal for various drive
field detunings. The influence of field $a$ detuning $\Delta_{g}$
is seen in Fig. \ref{fig:shapes}. A number of pronounced effects
comes about as this laser drives the Raman scattering and produces
ground-state atomic coherence $\rho_{gh}$. Initially the 2Ph signal grows exponentially. The corresponding
Raman gain coefficient is strongly dependent on drive field detuning
$\Delta_{g}$. 
The final effect is shortening of 2Ph pulses closer to resonance, as shown in Fig.~\hyperref[fig:shapes]{\ref*{fig:shapes}(b)}. The 4Ph pulse follows
the ground-state coherence $\rho_{gh}$ and 2Ph pulse as shown
in the previous section, however its maximum is somewhat delayed.
We attribute this effect to internal atom dynamics at high drive intensity
levels which is not captured by Eq.~\hyperref[eq:P2ndrho]{(\ref*{eq:P2ndrho})}. Energies
of pulses are also higher closer to resonance, although the two-photon
Raman pulse energy saturates due to absorption losses [see Fig.~\hyperref[fig:shapes]{\ref*{fig:shapes}(c)}].

Important insight is provided by calculating energy correlation between
the 2Ph and 4Ph light pulses, which fluctuate significantly from shot
to shot. Correlation is calculated as $C=\langle\Delta \mathcal{E}_{\mathrm{{2Ph}}}\Delta \mathcal{E}_{\mathrm{{4Ph}}}\rangle/\sqrt{\langle\Delta \mathcal{E}_{\mathrm{{2Ph}}}^{2}\rangle\langle\Delta \mathcal{E}_{\mathrm{{4Ph}}}^{2}\rangle}$,
where $\mathcal{E}_{\mathrm{{2Ph}}}$ and $\mathcal{E}_{\mathrm{{4Ph}}}$ are total energies
of 2Ph and 4Ph light pulses, respectively, in a single realization.
The averaging $\langle.\rangle$ is done over 500 realizations of
the process and results are plotted in Fig.~\hyperref[fig:shapes]{\ref*{fig:shapes}(d)}.
Strong correlations at various detunings reinforce the observation that indeed we are able
to couple 4Ph optical field to the ground-state coherence, since number
of photons in the 2Ph pulse is proportional to generated atomic coherence $|\rho_{gh}|^2$.
We attribute the drop in correlations close to resonance line to absorption
losses. Finally, we estimate the efficiency of conversion from the ground-state atomic coherence $\eta=\mathcal{E}_\mathrm{4Ph}/\mathcal{E}_\mathrm{2Ph}$ to the 4Ph field to be $5\times10^{-4}$. By comparing Eq.~(\ref{eq:P2ndrho}) with analogous expression for the 2Ph process given by Eq. \ref{eq:2ph}, we obtain $\eta\approx10^{-4}$ as well. This figure of merit could be improved by choosing different experimental geometries, or laser-cooling of the atomic ensemble.

\begin{figure}
\begin{centering}
\includegraphics[scale=1.15]{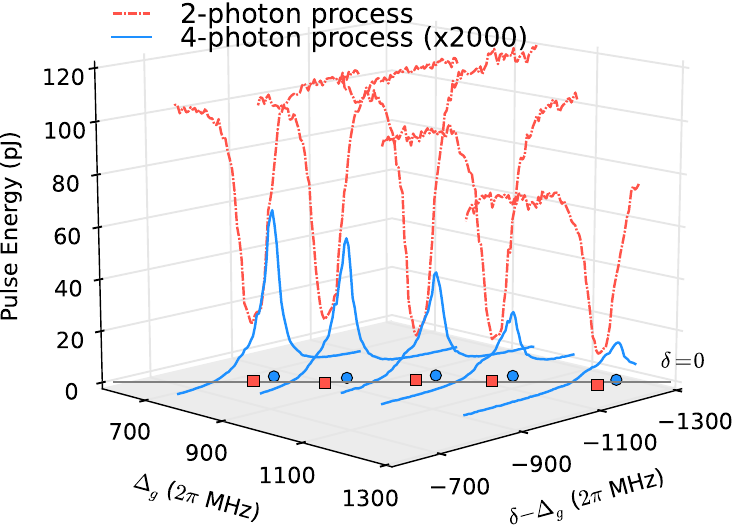} 
\par\end{centering}

\protect\caption{Experimental average pulse energies of 2Ph and 4Ph signals as a function
of field $a$ detuning $\Delta_{g}$ (measured from $F=1\rightarrow F'=1$
resonance) and field $b$ detuning $\delta-\Delta_{g}$ around the
two-photon resonance. As we change the field $a$ detuning both single
photon detuning $\Delta_{g}$ and the two-photon detuning $\delta$
change accordingly. Dots (squares) represent maxima (minima) of the 4Ph (2Ph)
signal, while $\delta=0$ line indicates the two-photon absorption resonance.}
\label{fig:detuning}
\end{figure}

To capture the physics in the vicinity of the two-photon resonance
we study 4Ph and 2Ph pulse energies as a function of detunings of
fields $a$ and $b$. First we scan the field $b$ detuning $\delta-\Delta_{g}$
across the two-photon resonance line (see Fig. \ref{fig:detuning}).
We observed strong suppression of 2Ph signal (minima are marked by squares in Fig. \ref{fig:detuning}) due to two-photon absorption
(TPA) in ladder configuration \cite{Moon2013}. As a consequence,
less atomic coherence $\rho_{gh}$ is generated and the 4Ph signal
is also reduced. In turn, the 4Ph shifted maxima position (marked by dots) are due to a trade-off
between TPA losses and 4WM efficiency, as the latter is highest at
the two-photon resonance (marked by $\delta=0$ line) according to Eq.~\hyperref[eq:P2ndrho]{(\ref*{eq:P2ndrho})}. The peak appears only on one side of the resonance due to the phase-matching condition being influenced by atomic dispersion \cite{Zibrov2002}. 
Additionally, we observe expected broadening of 60 MHz of two-photon resonance due to buffer gas collisions.
By changing the field $a$ detuning $\Delta_{g}$ we see expected
shifting of the two-photon resonance.  We checked that even at sub-optimal two-photon detuning $\delta$ the 2Ph and 4Ph signals are correlated, but the correlations become harder to measure as the 4Ph signals becomes very weak. 

\begin{figure}
\centering{}\includegraphics[scale=1.15]{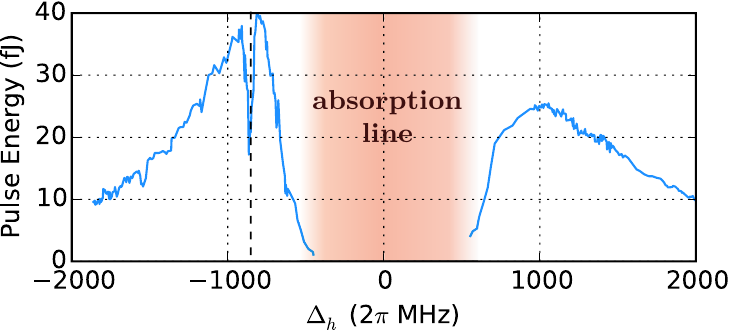}\protect\caption{Four-photon signal pulse energy as a function of the detuning $\Delta_{h}$
(measured from $F=2\rightarrow F'=2$ resonance) of field $c$ for
optimal conditions of other lasers ($\Delta_{g}/2\pi=900$~MHz and
$\delta/2\pi=-50$ MHz). Absorption line corresponds to the $F=2$
hyperfine component of the ground state, where the right side of the
plot is the red-detuned side. Drop at around $\Delta_{h}/2\pi=$-900
MHz corresponds exactly to 6.8 GHz detuning between the
two 780-nm lasers, or $\Delta_{g}+\Delta_{h}=0$.}
\label{fig:deltah}
\end{figure}

Contrary to the above, changing the detuning $\Delta_{h}$ of the
780-nm driving field $c$ has only mild effect on the 4Ph signal.
Fig.~\ref{fig:deltah} presents 4Ph signal pulse energy as a function
of $\Delta_{h}$ while other lasers were tuned for maximal signal
($\Delta_{g}/2\pi=900$ MHz and $\delta/2\pi=-50$ MHz). Since the
4Ph field frequency adapts to match the energy conservation for the
$|h\rangle\rightarrow|e_{3}\rangle\rightarrow|e_{2}\rangle$ two-photon
transition, the frequency of the driving field $c$ is not critical.
The laser must only be off-resonant, so the driving field is not absorbed
and does not disturb the ground-state coherence. We also observe	
a marked narrow drop in 4Ph process efficiency when the detuning between
fields $a$ and $c$ is exactly the ground-state hyperfine splitting, or equivalently $\Delta_{g}+\Delta_{h}=0$,
which is due to $\Lambda$ configuration two-photon resonance yielding strong interaction with the ground-state coherence. 

\section{Conclusions}\label{sec:concl}
The experiment we performed is a proof-of-principle
of a light atom interface that enables coupling of long-lived
ground-state atomic coherence and light resonant with transition between
excited states. The non-linear process we discussed is a novel type
of process with typical characteristics of both Raman scattering and
4WM. 
The observation of inverted-Y type nonlinear four-photon process involving ground-state coherence, performed in a very different regime in cold atoms, has been so far reported only in \cite{Ding2012}.
Here, we generated ground-state atomic coherence via the well known
two-photon process. We demonstrated ability to couple the
very same atomic coherence to optical field resonant with transition
between two excited states via a four-photon process. This was verified
by measuring high correlations between 2Ph and 4Ph fields, as well as frequency and
polarization characteristics of the four-photon process. 

We studied the behavior of pulse shapes as a function
of driving laser detunings. Among many results, we found that maximum
signal is achieved when lasers are detuned from the two-photon resonance
by approximately $\delta/2\pi=-50$ MHz, which is a trade-off
between TPA spoiling the generation of atomic coherence, and the efficiency
of the four-photon process. This results demonstrate that we are able
to control the 4WM process with ground-state coherence, which constitutes
a novel type of Raman scattering driven by three non-degenerate fields,
in analogy to hyper-Raman scattering where the scattering process is driven
by two degenerate fields.

Here we used light at 776 nm coupled to $5\mathrm{P}_{3/2}\rightarrow5\mathrm{D}_{5/2}$
transition. Using different states, such as $4\mathrm{D}_{3/2}$ as
the highest excited state, and $5\mathrm{P}_{1/2}$ and $5\mathrm{P}_{3/2}$
as two intermediate states, would enable coupling telecom light (at 1475.7 nm or 1529.3 nm). Such process could be used as a building
block for a telecom quantum repeater or memory \cite{Michelberger2015,Chaneliere2006,Zhang2016,Radnaev2010a}. By applying
external weak quantum field as the 4Ph field, the system may serve
as an atomic quantum memory, based on a highly non-linear process
as in \cite{DeOliveira2015}, but still linear in the input field
amplitude. It may also solve a variety of filtering problems \cite{Michelberger2015,Dabrowski2014},
since many similar configurations exists (e.g. with $5\mathrm{P}_{3/2}$,
$5\mathrm{D}_{3/2}$ and $5\mathrm{P}_{1/2}$ as intermediate states) in which all driving lasers operate at different wavelengths
than the signal. Even thought the 2Ph field was measured to be much stronger than the 4Ph field, we note that using a cold atomic ensemble would
offer selectivity in intermediate states of the process and small
detuning. Thanks to selection rules, exclusively the 4Ph process could be driven. This would be the requirement for generating pairs of photons and collective atomic excitations in the 4Ph process only.  Additionally, 4WM character of the
process enables engineering of phase-matching condition, namely changing
angles between incident driving beams to address different spin-wave excitations \cite{Chrapkiewicz2012},
to explore spatially-multimode capabilities of the system. In future
studies of the process we propose to address patterns 
unachievable
by typical Raman light-atom interface based on $\Lambda$ level configuration.

\begin{acknowledgments}
We acknowledge R. Chrapkiewicz, M. D\k{a}browski and J. Nunn for insightful discussions and  K. Banaszek and T. Stacewicz for their generous support.
This work was supported by Polish Ministry of Science and Higher Education ``Diamentowy Grant''
Project No. DI2013 011943 and National Science Center Grant
No. 2011/03/D/ST2/01941.
\end{acknowledgments}

\bibliography{bibliografia}

\end{document}